\documentclass[prl,twocolumn]{revtex4}
\usepackage{graphicx}
\begin{document}
\title{The Aharonov-Bohm-Casher ring-dot as a flux-tunable resonant tunneling diode}
\author{R. Citro$^{1,2}$ and F. Romeo$^{1}$}
\affiliation{$^{1}$ Dipartimento di Fisica ``E. R. Caianiello'',
C.N.I.S.M. and NANOMATES  (Research Centre for NANOMAterials and
nanoTEchnology), Universit{\`a} degli Studi di Salerno, Via S.
Allende, I-84081 Baronissi (Sa), Italy}\affiliation{$^{2}$
LPM$^2$C, Maison des Magisteres, CNRS, B.P. 166, F-38042 Grenoble,
France}

\begin{abstract}
A mesoscopic ring subject to the Rashba spin-orbit interaction and
sequentially coupled to an interacting quantum dot, in the
presence of Aharonov-Bohm flux, is proposed as a flux tunable
tunneling diode. The analysis of the conductance by means of the
nonequilibrium Green's function technique, shows an intrinsic
bistability at varying the Aharonov-Bohm flux when $2U> \pi
\Gamma$, $U$ being  the charging energy on the dot and $\Gamma$
the effective resonance width. The bistability properties are
discussed in connection with spin-switch effects and logical
storage device applications.
\end{abstract}


\maketitle In the last decade enormous attention has been devoted
towards control and engineering of spin degree of freedom in
nanostructures, usually referred to as
spintronics\cite{spintr_1,spintr_2}. Among the nanostructures of
major interest, quantum dots(QDs) provide information about the
fundamental physical phenomena in spin-dependent and strongly
interacting systems, such as the Kondo effect
\cite{kondo1,kondo2,kondo3,kondo4,kondo5,hewson_book}, the Coulomb
and the spin-blockade
effects\cite{blockade1,blockade2,blockade3,blockade4,blockade5},
spin-valve effect and tunneling magnetoresistance (TMR)
etc.\cite{tmr1,tmr2,tmr3,tmr4,tmr5}. Spin filter and
pumps\cite{filter1,filter2,pump,pump_fazio} have also been
proposed using QDs coupled to normal-metal leads.

The investigation of charging effects in tunneling transport
through small QDs has also opened up a large research field in the
last decades. In particular,  in the nonlinear regime, the current
voltage characteristic of a tunneling device containing a single
quantum dot exhibits step-like structures known as the Coulomb
staircase (for a review see \cite{review_1,review_2,review_3}). In
this context, double barrier resonant tunneling diodes (DBRTDs)
with InAs dots\cite{hysteresis_dot} represent a well known non
linear system that shows N (or in some cases Z-) shaped
current-voltage characteristics and intrinsic bistability.
Resonant tunneling diode devices may gain pratical applications in
microwave circuits. Their main feature is that they may become
bistable and thus useful as a logical storage device.

In the framework of quantum dot based devices, a system of great
interest is represented by a QD inserted in coherent ring
conductors. Here, suitable means for controlling spin are provided
by quantum interference effects under the influence of
electromagnetic potentials, known as
Aharonov-Bohm(AB)\cite{aharonov_bohm}  and
Aharonov-Casher(AC)\cite{aharonov_casher} effect. This possibility
has driven a wide interest in spin-dependent Aharonov-Bohm
physics, and the transmission properties of mesoscopic AB and AC
rings coupled to current leads have been studied under various
aspects (see e.g. \cite{romeo_cond}).

\begin{figure}[htbp]
\centering
\includegraphics[scale=.38]{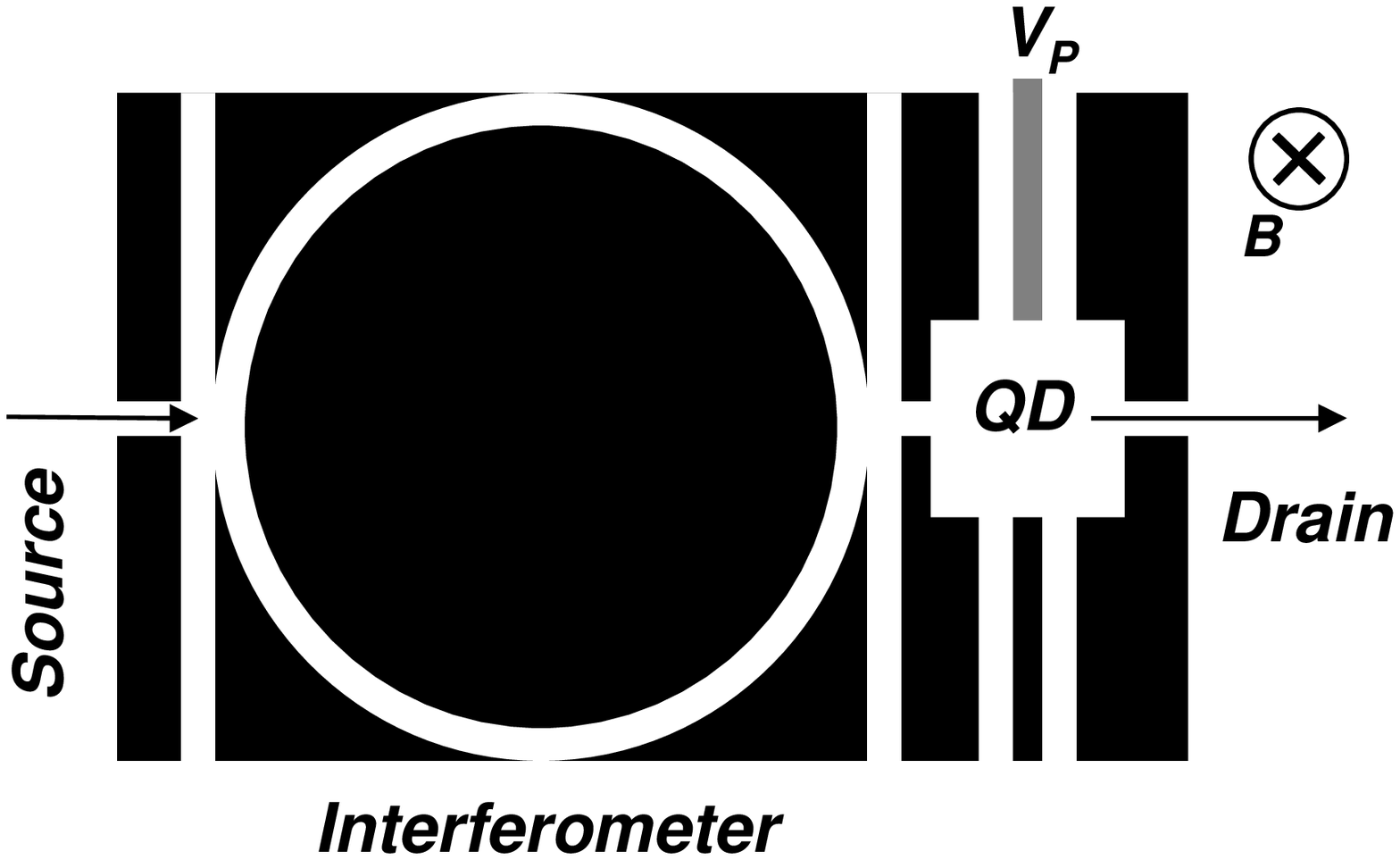}
\caption{An Aharonov-Bohm-Casher ring sequentially coupled to a
quantum dot (QD) (white square). The cross indicates the direction
of the external magnetic field orthogonal to the plane.}
\label{fig:abc-ring-dot-diode}
\end{figure}

In this paper we propose a ring-dot device with Rashba spin-orbit
interaction as a flux-tunable resonant tunneling diode.
 We consider a ring-dot
system coupled to two external leads (see
Fig.\ref{fig:abc-ring-dot-diode}) whose Hamiltonian in the local
spin frame is given by:
\begin{eqnarray}
&&\mathcal{H}=\sum_{\alpha=L,R}\mathcal{H}_\alpha+\mathcal{H}_d+\mathcal{H}_t\nonumber
\\
&&\sum_{\alpha=L,R}\mathcal{H}_\alpha+\mathcal{H}_d=\sum_{k,\sigma,\alpha\in
L,R}\varepsilon^{\alpha}_k c^{\dagger}_{k \sigma \alpha}c_{k
\sigma \alpha}+\sum_{\sigma}\varepsilon_{\sigma}d^{\dagger}_{\sigma}d_{\sigma}\nonumber\\
&&\mathcal{H}_t= \sum_{k, \sigma}\lbrack w c^{\dagger}_{k \sigma
R} d_{\sigma}+2u \cos(\varphi_{\sigma})d_{\sigma}^{\dagger}c_{k
\sigma L}+h.c.\rbrack,
\end{eqnarray}
where $\mathcal{H}_\alpha$ is the free electron Hamiltonian of the
leads ($\alpha=L,R$), $c_{k\sigma \alpha}(c_{k\sigma
\alpha}^\dagger)$, being the annihilation(creation) operator of
the conduction electrons and
$\varepsilon^{\alpha}_k=\tilde{\varepsilon}^{\alpha}_k-\mu_\alpha$
with $\mu_\alpha$ the chemical potential; $\mathcal{H}_d$ is the
Hamiltonian of dot where $d_\sigma(d_\sigma^\dagger)$ is the
annihilation(creation) operator of the electrons on the dot.
Within the Hartree-Fock approximation for the Coulomb interaction
$U$ on the level dot\cite{noteHF},
$\varepsilon_\sigma=\epsilon_0-\mu+U\langle n_{\bar{\sigma
}}\rangle$, where $\langle n_\sigma \rangle=\langle
d^{\dagger}_{\sigma}d_{\sigma}\rangle$ and $\epsilon_0-\mu$ is
controlled via a gate voltage. Finally, $\mathcal{H}_t$ describes
the tunneling between the dot and the leads, where in the last
term $\varphi_{\sigma}$ ($\sigma=\pm$) is the effective flux
enclosed in the ring. In particular,
$\varphi_{\sigma}=\pi(\Phi_{AB}+\sigma\Phi_{R})$, $\Phi_{AB}$
being the AB flux induced by a perpendicular magnetic field and
$\Phi_R$ being the effective flux induced by the Rashba spin-orbit
interaction\cite{aharonov_casher}. The Aharonov-Casher flux is
explicitly given by $\Phi_{R}=\sqrt{\beta^2+1}$, where
$\beta=2\delta m^\star/\hbar^2$ is the dimensionless spin-orbit
interaction, $m^\star$ is the effective mass of the carriers and
the parameter $\delta$ is related to the average electric field
along the direction orthogonal to the plane of the
ring\cite{romeo_cond}. This is assumed to be a tunable quantity.
For an InGaAs-based two-dimensional electron gas, $\delta$ can be
controlled by a gate voltage with typical values in the range
$(0.5\div 2.0)\times 10^{-11}$eVm\cite{param1,param2}. An external
bias voltage $V$ drives the system away from equilibrium thus
imposing a chemical potential imbalance between the left (L) and
the right (R) leads, $\mu_L=0$ and $\mu_R=-eV$, where $e$ is the
absolute value of the electron charge. The dot level can be tuned
by means of a local gate voltage $V_p$ almost independently from
the voltage drop between the external leads.
The tunneling amplitudes $w$ and $u$, which in general may depend
on the spin and electron momentum, are assumed to be constant. To
calculate the current we apply the Green's function (GF) approach
within the Keldysh formalism\cite{meir_gf,jauho_gf}. The
well-known equation for the stationary current reads:
\begin{eqnarray}
\label{eq:curr}&&I=\frac{ie}{2 h} \int d \varepsilon Tr \{
[\mathbf{\Gamma}^L(\varepsilon)-\mathbf{\Gamma}^R(\varepsilon)]\mathbf{G}^{<}(\varepsilon)+\\\nonumber
&&+[f_L(\varepsilon)\mathbf{\Gamma}^L(\varepsilon)-f_R(\varepsilon)\mathbf{\Gamma}^R(\varepsilon)](\mathbf{G}^r(\varepsilon)-\mathbf{G}^a(\varepsilon))\},
\end{eqnarray}
where $\mathbf{G}^r,\mathbf{G}^a$ and $\mathbf{G}^<$ are the
retarded, advanced and lesser dot Green's function, respectively,
$f_\alpha(\omega)$ is left/right Fermi-Dirac distribution function
and  $\mathbf{\Gamma}^{L,R}$ are the level width (the boldface
notation indicates matrices in the spin basis). To calculate these
GFs we apply the equation of motion
technique\cite{jauho_review,flensberg_book} and the retarded
Green's function is given by:
\begin{equation}
\label{eq:RGF}
G^r_{\sigma\sigma'}(\omega)=[g^r(\omega)^{-1}-\Sigma^{r}(\omega)]^{-1}_{\sigma\sigma'},
\end{equation}
where  $\Sigma^{r}_{\sigma\sigma'}(\omega)$ is the tunneling
self-energy:
\begin{eqnarray}
\label{eq:RSE}
\Sigma^{r}_{\sigma\sigma'}(\omega)&=&\sum_{k,\alpha=R}
\gamma^2|u|^2[g^{r}_{kR}]_{\sigma\sigma'}(\omega)+\\\nonumber
&&+\sum_{k,\alpha=L}4|u|^2\cos(\varphi_\sigma)\cos(\varphi_{\sigma'})[g^{r}_{kL}]_{\sigma\sigma'}(\omega).
\end{eqnarray}
Here we have introduced the retarded Green's functions of the
leads
$[g^{r}_{k\alpha}]_{\sigma\sigma'}=\delta_{\sigma\sigma'}(\omega+i0^{+}-\varepsilon_k^{\alpha})^{-1}$
and the retarded Green's function of the dot without coupling to
the leads
$g^r_{\sigma\sigma'}=\delta_{\sigma\sigma'}(\omega+i0^{+}-\varepsilon_\sigma)^{-1}$.
The latter depends explicitly on the occupation of the dot which
is determined self-consistently by $\langle n_\sigma\rangle=\int
\frac{d\omega}{2\pi i}G^{<}_{\sigma\sigma}(\omega)$. By evaluating
(\ref{eq:RSE}), the retarded Green's function of the dot can be
written as:
\begin{equation}
\label{eq:gr}
G^r_{\sigma\sigma'}(\omega)=\delta_{\sigma\sigma'}\{\omega-\varepsilon_\sigma+i\pi|u|^2[\rho_R\gamma^2+4\rho_L\cos^2(\varphi_\sigma)]\}^{-1},
\end{equation}
where the  parameter $\gamma$ is defined as $w=\gamma u $, while
the density of states $\rho_{L/R}(\omega)$ is given by
$\sum_{k,\alpha}\left(\omega+i0^{+}-\varepsilon^{\alpha}_k\right)^{-1}$.
The correlation function $G^{<}_{\sigma\sigma'}$ is given by the
Keldysh equation
$\mathbf{G}^{<}=\mathbf{G}^r\mathbf{\Sigma}^<\mathbf{G}^a$, where
the advanced Green's function is
$\mathbf{G}^a(\omega)=[\mathbf{G}^r(\omega)]^\dagger$, while
$\mathbf{\Sigma}^{<}(\omega)=i[f_R(\omega)\mathbf{\Gamma}^R+f_L(\omega)\mathbf{\Gamma}^L]$.
In the wide band limit (WBL), the level widths
$\Gamma^{\alpha}_{\sigma\sigma'}$  are taken energy independent
and they are given by
$\Gamma^{L}_{\sigma\sigma'}=8\pi\rho_L|u|^2\cos^2(\varphi_\sigma)\delta_{\sigma\sigma'}$
and
$\Gamma^{R}_{\sigma\sigma'}=2\pi\rho_R|u|^2\gamma^2\delta_{\sigma\sigma'}$,
where $\rho_{L/R}\sim \rho$ at the Fermi level.

The final expression of the lesser Green's function is:
\begin{equation}
\label{eq:gless}
G^{<}_{\sigma\sigma'}=i\delta_{\sigma\sigma'}\frac{f_R(\omega)\Gamma_{\sigma\sigma}^R+f_L(\omega)\Gamma_{\sigma\sigma}^L}
{(\omega-\varepsilon_\sigma)^2+(\Gamma_\sigma/2)^2}
\end{equation}
where the quantity
$\Gamma_\sigma=\Gamma^{L}_{\sigma\sigma}+\Gamma^{R}_{\sigma\sigma}$
has been introduced. By using (\ref{eq:gr}) and (\ref{eq:gless})
and the definition (\ref{eq:curr}), the current per spin channel
flowing through the system is given by:
\begin{equation}
\label{eq:curr_int} I_\sigma=(e/\hbar)\int
\frac{d\varepsilon}{2\pi}\frac{\Gamma^L_{\sigma\sigma
}\Gamma^R_{\sigma\sigma
}}{(\varepsilon-\varepsilon_\sigma)^2+(\Gamma_\sigma/2)^2}(f_L(\varepsilon)-f_R(\varepsilon)).
\end{equation}

After evaluating the integral(\ref{eq:curr_int}), the differential
conductance per spin channel (in units of $e^2/h$) in the linear
response regime\cite{note} and the occupation on the dot are
explicitly given by:
\begin{eqnarray}
\label{eq:conductance-ring-diode}
G_{\sigma}&=&\frac{\gamma^2\cos^{2}(\varphi_\sigma)}{(\Delta-U \langle n_{\bar{\sigma}}\rangle)^2+(\gamma^2/4+\cos^2(\varphi_\sigma))^2}\\
\label{eq:conductance-ring-diode_n} \langle n_\sigma\rangle &=&
\frac{1}{2}+\frac{1}{\pi}\arctan \Bigl[\frac{\Delta-U\langle
n_{\bar{\sigma}}\rangle}
{\gamma^2/4+\cos^2(\varphi_\sigma)}\Bigl],
\end{eqnarray}
where $\Delta=e(V+V_p)-\epsilon_0$ and the energy is measured in
units of $4\pi \rho |u|^2$. The quantity $\Delta$ can be tuned by
means of the gate voltage $V_p$ on the dot and the bias voltage
$V$. It is worth to notice that the spin up/down occupations are
now coupled due to the combined effect of the Coulomb interaction
and the spin-orbit interaction.

\begin{figure}[htbp]
\centering
\includegraphics[scale=.38]{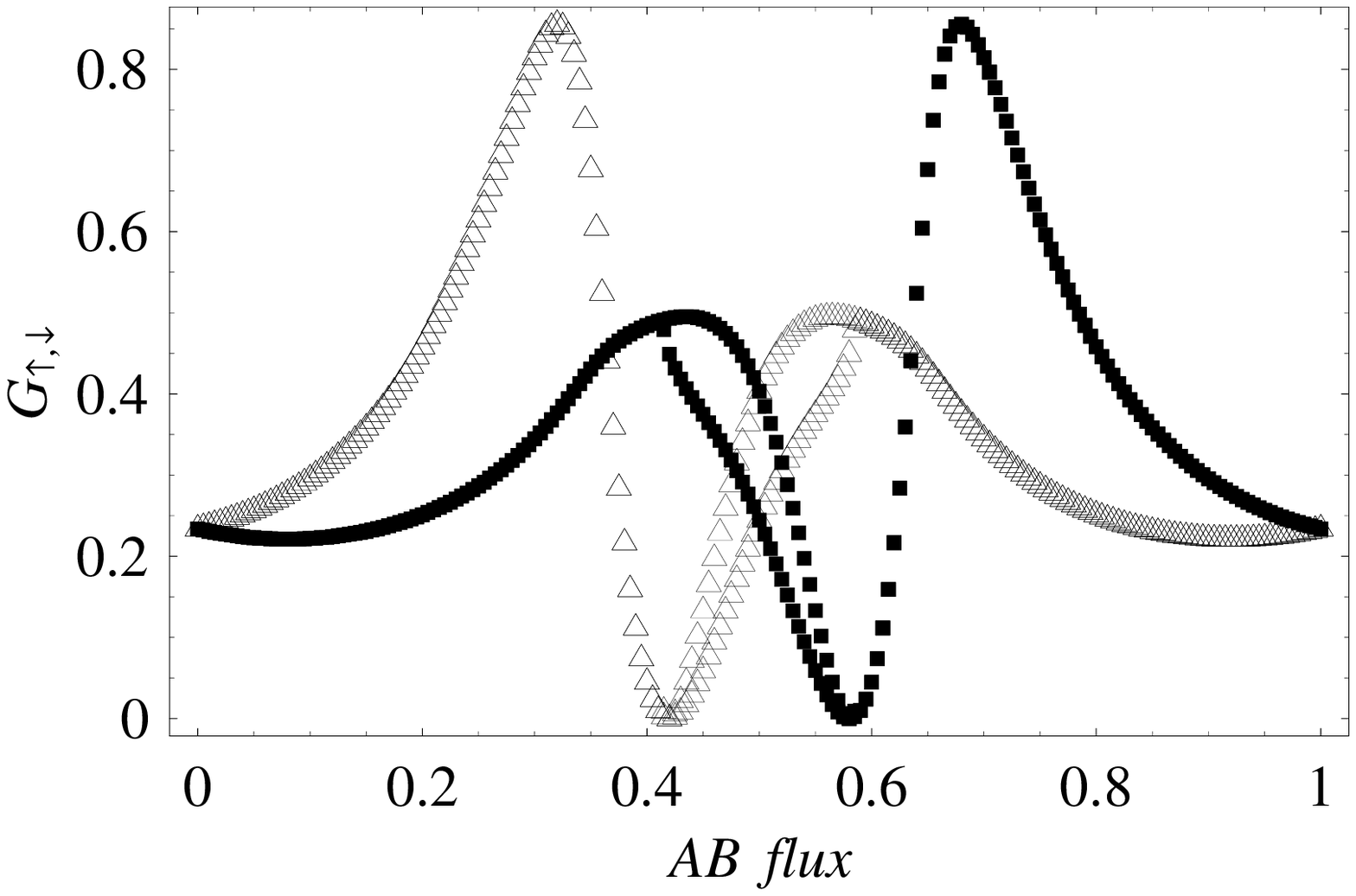}\\
\vspace{1.5truecm}
\includegraphics[scale=.38]{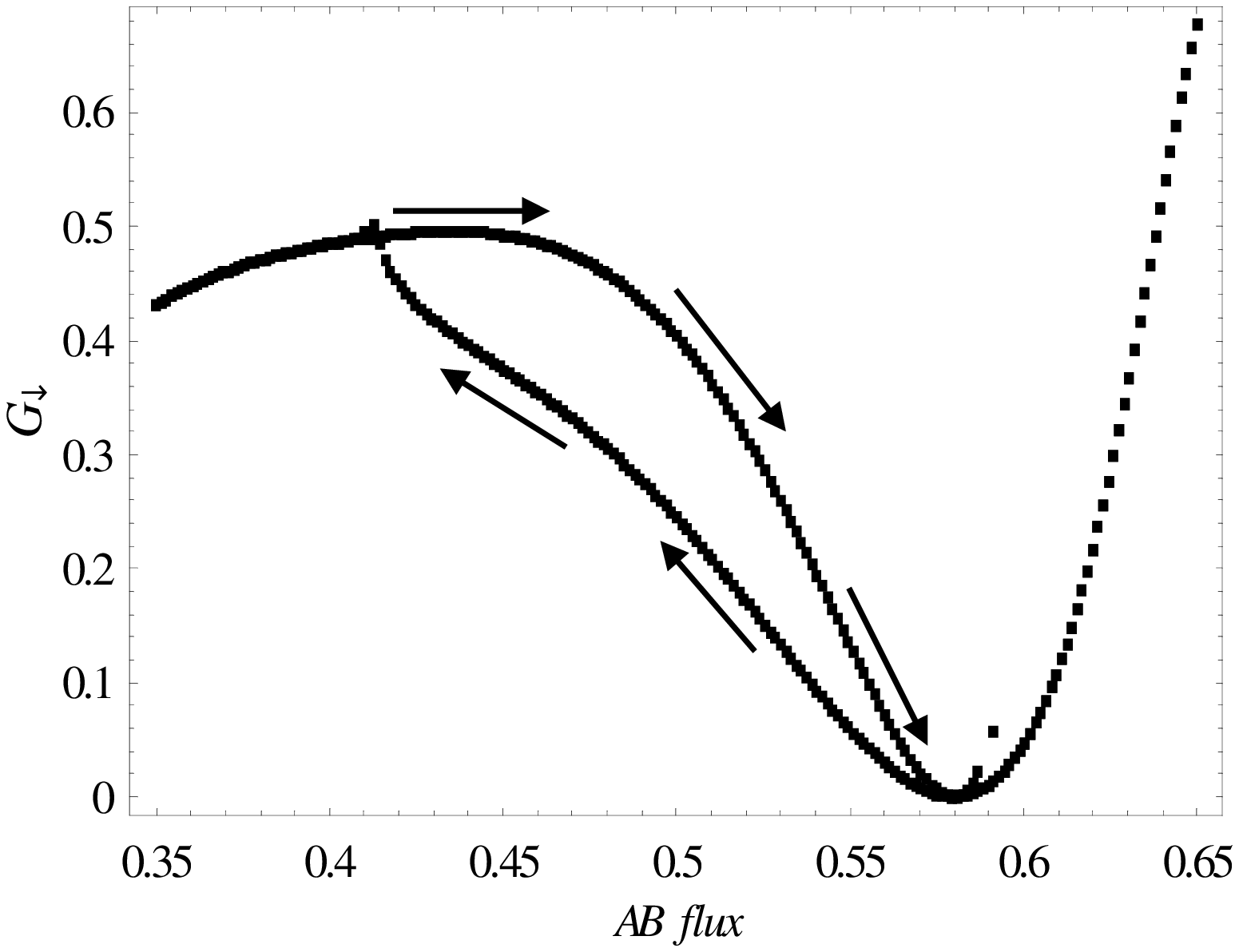}
\\
\caption{\textbf{Upper panel}: Conductance curves $G_{\uparrow}$
(triangle) and $G_{\downarrow}$ (box) as a function of the AB flux
$\Phi_{AB}$ for  $U=0.6$, $\Delta=0.25$,
$\Phi_R=0.08$,$\gamma=0.5$. A spin-filter effect is obtained close
to $\Phi_{AB}=0.42$ or $\Phi_{AB}=0.58$. \textbf{Lower panel}:
Hysteretic behavior of $G_{\downarrow}$ (box) close to
$\Phi_{AB}=0.42$ for the same parameters as above. The conductance
$G_{\uparrow}$ (triangle) is strongly suppressed close to
$\Phi_{AB}=0.42$}. \label{fig:cond-hyst-ring-dot-diode}
\end{figure}

In the upper panel of Fig.\ref{fig:cond-hyst-ring-dot-diode} the
conductance curves $G_\uparrow$ and $G_\downarrow$ are shown as a
function of the AB flux by fixing the other model parameters as
$U=0.6$, $\Delta=0.25$, $\Phi_R=0.08$ and $\gamma=0.5$ (energies
are in units of $4\pi \rho |u|^2$). In the vicinity of a zero of
the conductance of a given spin channel there corresponds a finite
value of the conductance in the other spin channel, giving rise to
a remarkable magneto-resistance effect. This behavior is
originated by an intrinsic bistability as shown in the lower panel
of Fig.\ref{fig:cond-hyst-ring-dot-diode} where the hysteresis
loop in the AB flux is shown. Since from
(\ref{eq:conductance-ring-diode_n}) the variation of the AB flux
would correspond to a variation of the occupation of the dot, the
intrinsic bistability effect could be understood as a charging
effect. It arises since the charge trapped in the resonant dot
state changes the potential profile through the double well
potential, and therefore modifies the energy of the resonant
state.  Each time an electron jumps into the dot, the dot
potential is lift up due to the charging effect, until the
electron tunnels through. The potential is lowered again and it
picks up an electron until bias reach a certain condition where
the dot no longer picks up an electron after emitting the
previously captured one. Thus there is a range of AB fluxes over
which the resonance can carry current. From the analytical
expression (\ref{eq:conductance-ring-diode_n}) one may deduce a
condition for the bistability in the following way. Expressing
(\ref{eq:conductance-ring-diode_n}) only in terms of $\langle
n_\sigma \rangle$ and employing the lowest order expansion in the
quantity $U\langle n_{\sigma} \rangle$, the equation for the dot
occupation becomes:
\begin{equation}\label{eq:cond}
\tan  (\pi (\langle n_\sigma
\rangle-\frac{1}{2}))=T_\sigma-\beta_\sigma \langle n_\sigma
\rangle,
\end{equation}
where
$T_{\sigma}=A_{\sigma}+B_{\sigma}\arctan(q_{\bar{\sigma}}\Delta)$,
while $\beta_{\sigma}=\frac{U
B_{\sigma}q_{\bar{\sigma}}}{1+(\Delta q_{\sigma})^2}$, with
$q_{\sigma}=(\frac{\gamma^2}{4}+\cos^{2}(\varphi_{\sigma}))^{-1}$,
$A_{\sigma}=q_{\sigma}(\Delta-U/2)$ and
$B_{\sigma}=-Uq_{\sigma}/\pi$. It can be analytically proved that
Eq.(\ref{eq:cond}) admits more then one solution when $\pi \Gamma
\leq 2U$ where $\Gamma=\sqrt{\Gamma_\sigma \Gamma_{\bar{\sigma}}}$
is an effective level width. The obtained relation is similar to
the one for a resonant tunneling diode (RTD) where an intrinsic
bistability has been demonstrated both theoretically and
experimentally\cite{rahman_diode,diode_exp}. In terms of the
charging energy ($U=e^2/2C$) the condition to have bistability in
our case may be written as:
\begin{equation}\label{eq:hyst-cond}
\frac{e Q_m}{C}>\pi\Gamma,
\end{equation}
where the maximum charge density on the dot $Q_m$ has been
introduced. Thus the change in energy of the resonance caused by
the stored charge must exceed the width of the resonance for the
bistability to be seen. This is a physically appealing condition
for a device to show intrinsic bistability. Furthermore, for a
fixed strength of the charging energy $U$, the hysteresis
condition can be more easily fulfilled for the values of
$\varphi_{\sigma}$ such that $\cos(\varphi_{\sigma})\sim 0$ and
for a sufficiently small value of $\gamma$. As already shown in
the case of the RTD by M. Rahman and J. H. Davies
\cite{rahman_diode}, the bistability condition is favoured by
making the exit barrier more opaque (i.e. for small values of
$\gamma$ in our case) even though the current flowing through the
system decreases. The main difference from standard RTD is that in
the Rashba ring tunneling is governed by a flux dependent barrier
with effective strength $2u\cos(\varphi_{\sigma})$ and acting
selectively on the different spin channels. We would also like to
point out that the physics of the bistability effect is similar to
that of the appearance of a magnetic phase for localized states in
metals at varying the local moment level\cite{doniach_book}.

\begin{figure}[htp]
\centering
\includegraphics[scale=.42]{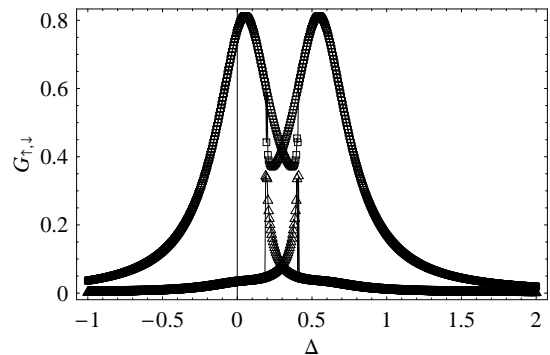}
\caption{The Conductance $G_{\uparrow}$ (triangle) and
$G_{\downarrow}$ (box) as a function of $\Delta$ for $U=0.6$,
$\Phi_{AB}=0.45$, $\Phi_R=0.08$,
$\gamma=0.5$.}\label{fig:cond-delta-ring-diode}
\end{figure}

In Fig.\ref{fig:cond-delta-ring-diode}  the conductance
$G_{\uparrow,\downarrow}$ is shown as a function of $\Delta$ by
fixing the other parameters as $U=0.6$, $\Phi_{AB}=0.45$,
$\Phi_R=0.08$,$\gamma=0.5$. In the figure, charging energy peaks
are evident and strong asymmetric resonances (Fano-like) are
observed in the conductance $G_{\uparrow}$ close to $\Delta=0.25$.
Such resonances, characterized by a very long life-time, are
responsible for creating the condition to have bistability.
On the other hand, we have been checking that increasing the
parameter $\gamma$ while fixing the other parameters as done in
Fig.\ref{fig:cond-hyst-ring-dot-diode}, the hysteresis is lost.
Indeed, an increase of $\gamma$ produces an increasing of the
second term in Eq.(\ref{eq:hyst-cond}) making the inequality
false.
%

In conclusion, we studied the conductance of a mesoscopic ring
sequentially coupled to an interacting quantum dot in the presence
of Rashba spin-orbit interaction and an Aharonov-Bohm flux. By
treating the Coulomb interaction on the dot within a
self-consistent mean field approximation and by exploiting the
non-equilibrium Green's functions technique, we demonstrated that
the system behaves as a flux-tunable resonant tunneling diode. In
analogy to the case of the standard resonant tunneling diode it
has been shown, both analytically and numerically, that an
hysteretic behavior is observed in the conductance of a given spin
channel at varying the flux. The bistability properties are
characterized by a spin-switch effect in which the conductance of
one spin channel is totally suppressed while the other is close to
the maximum value. When the hysteresis condition is not fulfilled
the system behaves like a traditional spin interference device as
the one proposed in the pioneering work by J. Nitta et
al.\cite{nitta-sp-int-dev}. The proposed device can be easily
realized by e.g. InAs quantum dots embedded in a
Aharonov-Bohm-Casher ring\cite{ring_exp} and potentially employed
as high speed switching two-terminal device or a useful logical
storage device\cite{storage}.

The authors acknowledge Prof. M. Marinaro for helpful and
enlightening discussions. R.C. acknowledges support by the
European community under the Marie Curie Program.


\begin{thebibliography}{99}
\bibitem{spintr_1}G. A. Prinz, Science {\bf 282}, 1660 (1998).
\bibitem{spintr_2} S. A. Wolf, D. D. Awschalom, R. A. Buhrman, J. M. Daughton, S.
von Moln\'{a}r, M. L. Roukes, A. Y. Chtchelkanova, and D. M.
Treger, Science {\bf 294}, 1488 (2001).
\bibitem{kondo1}R. Swirkowicz, M.
Wilczynski, M. Wawrzyniak, and J. Barnas, Phys. Rev. B
\textbf{73}, 193312 (2006). \bibitem{kondo2} J. Martinek, M.
Sindel, L. Borda, J. Barnas, R. Bulla, J. K\"{o}nig, G. Sch\"{o}n,
S. Maekawa, and J. von Delft, Phys. Rev. B \textbf{72}, 121302 R
(2005).
\bibitem{kondo3}Y.
Utsumi, J. Martinek, G. Sch\"{o}n, H. Imamura, and S. Maekawa,
Phys. Rev. B \textbf{71}, 245116 (2005).
\bibitem{kondo4}J. Martinek, Y. Utsumi,
H. Imamura, J. Barnas, S. Maekawa, J. K\"{o}nig, and G. Sch\"{o}n,
Phys. Rev. Lett. \textbf{91}, 127203 (2003).
\bibitem{kondo5} P. Zhang, Q.-K.
Xue, Y. Wang, and X. C. Xie, Phys. Rev. Lett. \textbf{89}, 286803
(2002).
\bibitem{hewson_book} A. C. Hewson, \textit{The Kondo Problem for Heavy Fermions}
,Cambridge University Press, Cambridge, (1993).
\bibitem{blockade1} F. Elste and C. Timm, Phys. Rev. B
\textbf{73}, 235305 (2006). \bibitem{blockade2} I. Weymann, J.
Barnas, J. K\"{o}nig, J. Martinek, and G. Sch\"{o}n, Phys. Rev. B
\textbf{72}, 113301 (2005).
\bibitem{blockade3} A. Cottet, W. Belzig, and C. Bruder, Phys. Rev. Lett. \textbf{92},
206801 (2004). \bibitem{blockade4} J. Barnas and A. Fert, Phys.
Rev. Lett. \textbf{80}, 1058 (1998).
\bibitem{blockade5} S. Takahashi and
S. Maekawa, Phys. Rev. Lett. \textbf{80}, 1758 (1998).
\bibitem{tmr1} J.
Varalda, A. J. A. de Oliveira, D. H. Mosca, J.-M. George, M.
Eddrief, M. Marangolo, and V. H. Etgens, Phys. Rev. B \textbf{72},
081302(R) (2005). \bibitem{tmr2} I. Weymann, J. K\"{o}nig, J.
Martinek, J. Barnas, and G. Sch\"{o}n, Phys. Rev. B \textbf{72},
115334 (2005).
\bibitem{tmr3} R. L\'{o}pez and D. S\'{a}nchez, Phys. Rev.
Lett. \textbf{90}, 116602 (2003). \bibitem{tmr4} J. K\"{o}nig and
J. Martinek, Phys. Rev. Lett. \textbf{90}, 166602 (2003).
\bibitem{tmr5} Braun, J. K\"{o}nig, and J. Martinek, Phys. Rev. B
\textbf{70}, 195345 (2004).
\bibitem{filter1}P. Recher, E. V. Sukhorukov, and D. Loss, Phys. Rev.
Lett. \textbf{85}, 1962 (2000).
\bibitem{filter2} H.-A. Engel and D. Loss,
Phys. Rev. B \textbf{65}, 195321 (2002).
\bibitem{pump} E. Cota, R. Aguado,
and G. Platero, Phys. Rev. Lett. \textbf{94}, 107202 (2005).
\bibitem{pump_fazio}  J. Splettstoesser, M. Governale, J. K\"{o}nig,
R.Fazio, Phys. Rev. B \textbf{74}, 085305 (2006).
\bibitem{review_1} {\it Single Charge Tunneling}, Vol. 294
of NATO Advanced Study Institute, Series B: Physics, edited by H.
Grabert and M.H. Devoret ~Plenum, New York, (1992).
 \bibitem{review_2} U. Meirav and
E.B. Foxman, Semicond. Sci. Technol. {\bf 10}, 255 (1995).
 \bibitem{review_3} L.P.
Kouwenhoven, C.M. Marcus, P.L. McEuen, S. Tarucha, R. M.
Westervelt, and N.S. Wingreen, in {\it Mesoscopic Electron
Transport}, edited by L.L. Sohn, L.P. Kouwenhoven, and G.
Sch\"{o}n ~Kluwer Academic Publishers, Dordrecht, (1997).
\bibitem{hysteresis_dot} R. Tsu and L. Esaki, Appl. Phys. Lett. \textbf{22}, 562
(1973); V.J. Goldman, D.C. Tsui, and J.E. Cunningham, Phys. Rev.
Lett. \textbf{58}, 1256 (1987); A.D. Martin, M.L.F. Lerch, P.E.
Simmonds, and L. Eaves, Appl. Phys. Lett. \textbf{64}, 1248
(1994).
\bibitem{aharonov_bohm}Y. Aharonov and D. Bohm, Phys. Rev. \textbf{115}, 485 (1959).
\bibitem{aharonov_casher} Y. Aharonov and A. Casher, Phys. Rev. Lett. \textbf{53}, 319 (1984).
\bibitem{romeo_cond} R. Citro, F. Romeo, M. Marinaro, Phys. Rev. B {\bf 74}, 115329
(2006); R. Citro and F. Romeo, Phys. Rev. B {\bf 71}, 073306
(2007).
\bibitem{noteHF} The mean-field approximation neglects the dynamical
properties of the local field (Kondo effect) and gives reliable
results away from the Coulomb blockade regime.
\bibitem{param1} J. Nitta, T. Akazaki, H. Takayanagi, and T. Enoki, Phys. Rev.
Lett. {\bf 78}, 1335 (1997).
\bibitem{param2} D. Grundler, Phys. Rev. Lett. {\bf 84}, 6074 (2000).
\bibitem{meir_gf} Y. Meir and N.S. Wingreen, Phys. Rev. Lett. {\bf
68}, 2512 (1992).
\bibitem{jauho_gf} A.P. Jauho and N.S. Wingreen, and Y. Meir,
Phys. Rev. B, {\bf 50}, 5528 (1994).
\bibitem{jauho_review} H. Haug and A. P. Jauho, {\it Quantum Kinetics in Trasport and Optics of
Semiconductors}, (Springer, New York, 1996).
\bibitem{flensberg_book} H. Bruus and K. Flensberg, \textit{Many-Body Quantum Theory in Condensed-Matter Physics: An Introduction}
(Oxford University Press, Oxford, 2004).
\bibitem{note} It is assumed that the dot level is on resonance
within the conduction window between $\mu_L$ and $\mu_R$ for
positive bias.
\bibitem{diode_exp} K. Yoh, H. Kazama, Y. Kitashou, and T. Nakano,
Phys. Stat. Sol. (b) \textbf{204}, 378 (1997).
\bibitem{rahman_diode} M. Rahman and J.H. Davies, Semicond. Sci.
Technol. {\bf 5}, 168 (1990).
\bibitem{doniach_book} S. Doniach, E.H. Sondheimer, {\it Green's
functions for Solid State Physicists}, Imperial College Press,
London (2004).
\bibitem{nitta-sp-int-dev} J. Nitta et al., Appl. Phys. Lett. \textbf{75},
695 (1999).
\bibitem{ring_exp}N. T. Bagraev, N. G. Galkin, W. Gehlhoff, L. E. Klyachkin, A. M. Malyarenko and I. A. Shelykh,
Journal of Physics: Conference Series \textbf{61}, 56-60 (2007).
\bibitem{storage} K. Nakazato, R.J. Blaikie and H. Ahmed, J. Appl.
Phys. {\bf 75}, 5123 (1994).
\end{thebibliography}
\end{document}